\documentclass{ptephy}

\usepackage{boites}
\usepackage{setspace}
\usepackage{graphicx}

\begin{document}

\title{Performance of a Remotely Located Muon Radiography System to Identify the Inner Structure of a Nuclear Plant}

\author{\name{H. Fujii}{1}, \name{K. Hara}{2,\ast}, \name{S. Hashimoto}{2}\thanks{Now at NEDO, 1310 Omiya, Saiwai-ku, Kawasaki, Kanagawa 212-8554, Japan}, 
\name{F. Ito}{2}, \name{H.~Kakuno}{3}, \name{S.H.~Kim}{2}, \name{M. Kochiyama}{1}
\name{K.~Nagamine}{1}, \name{A.~Suzuki}{1}, \name{Y. Takada}{2}
\name{Y.~Takahashi}{2}\thanks{Now at JAXA, 3-1-1 Yoshinodai, Chuo-ku, Sagamihara, Kanagawa 252-5210, Japan}, 
\name{F.~Takasaki}{1}, \name{S.~Yamashita}{4}
}

\address{\affil{1}{KEK, 1-1 Oho, Tsukuba, Ibaraki 305-0801, Japan}
\affil{2}{Univ. of Tsukuba, 1-1-1 Tennodai, Tsukuba, Ibaraki 305-8571, Japan}
\affil{3}{Tokyo Metropolitan Univ., 1-1 Minami-Osawa, Hachioji, Tokyo 192-0397, Japan}
\affil{4}{Univ. of Tokyo, 7-3-1 Hongo, Bunkyo-ku, Tokyo 113-0033, Japan}
\email{hara@hep.px.tsukuba.ac.jp}}

\begin{abstract}%
The performance of a muon radiography system designed
to image the inner structure of a nuclear plant located at a distance of 64~m was evaluated.
We concluded absence of the fuel in the pressure vessel during the measurement period 
and succeeded in profiling the fuel material placed in the storage pool. 
The obtained data also demonstrated the sensitivity of the system to water level changes in the reactor well 
and the dryer-separator pool.
It is expected that the system could reconstruct a 2~m cubic fuel object easily. 
By operating multiple systems, typically four identical systems, 
viewing the reactor from different directions simultaneously, detection of a 1~m cubic object should
also be achievable within a few month period.
\end{abstract}

\subjectindex{xxxx, xxx}

\maketitle

\section{Introduction}

Soon after Fukushima Daiichi reactor accident, one of the authors (K.N.) proposed that it would be possible
to inspect damaged reactors using cosmic rays. 
Some pioneering works\cite{n0,n1,n2,n3,n4} have already used cosmic muons for the study of the inner structure 
of big objects such as volcanoes and hot melting furnaces: 
this method is often referred to as Muon Imaging or Muon Radiography. 
By accumulating millions of muon tracks traversing through such heavy and big objects, 
one can image their inner structure, much like medical X-ray imaging. 

A team was formed by the people of KEK, the University of Tsukuba, and the University of Tokyo to consider 
the possible applications of muon imaging technology to troubled nuclear reactors. 
It became immediately clear that the situation surrounding the reactors, for example the radiation level, 
would not allow an individual to bring the detector system inside the reactor building. 
The radiation levels just outside of the Fukushima Daiichi building were so high 
that even some protection of the detector against radiation needed to be carefully considered. 
The accuracy of the track position measurement was re-evaluated so that the detector system could provide a clear 
image of the inner structure even if placed outside of the reactor building.
An attempt to image the fuel in a reactor using emulsion 
plates located
directly underneath the reactor has been reported \cite{emulsion}. 
Setting a detector at the same place would be difficult in view of accessibility required for data collection. 
In light of these conditions, 
the KEK-Tsukuba team built a detector system using finely segmented hodoscopes consisting of 1 cm (wide) by 100 cm (long) scintillator bars. 
Wavelength shifting fibers were embedded in the bars and coupled with MPPCs (Multi-pixel photon counters or Silicon PMTs) for readout. The signal readout and trigger processing was performed using an FPGA (Field Programmable Gate Array) based system. 
These detector technologies, each of which had been developed originally for various individual
physics experiment, were integrated into a muon radiography system that could be adoptable to evaluate, 
otherwise very difficult, the inner structure of the huge nuclear plant.    

In order to evaluate the performance of the detector, we carried out a test measurement by imaging a GM MK-II type nuclear reactor at the 
Japan Atomic Power Company (the JAPC). 
The system was installed in March 2012 about 64 m away from the reactor center, and this 
report describes the design of the detector system, operational experience, and outcome of the test measurements carried out at the JAPC.

\section{Detector System}

\subsection{Overview of the detector}

\begin{figure}[htbp]
\begin{center}
\includegraphics[width=15.0cm, bb=0 0 848 377]{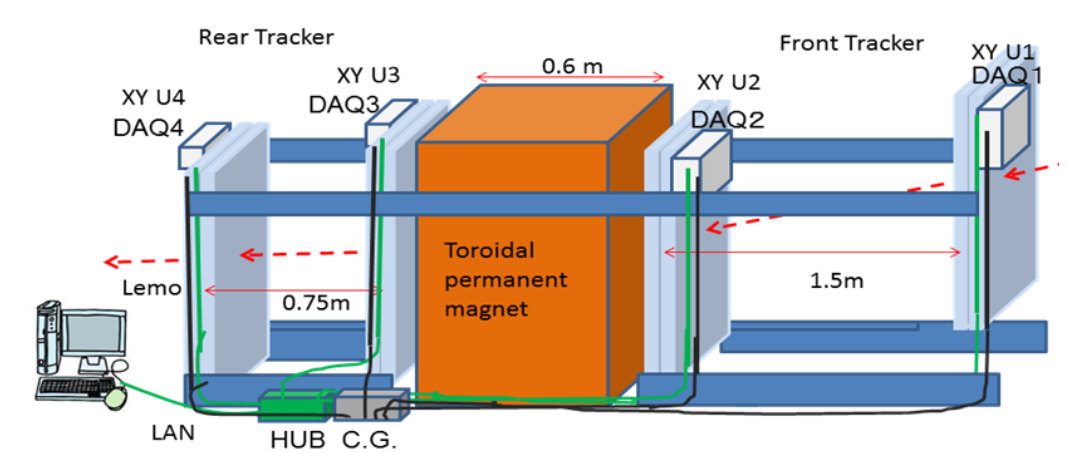}
\end{center}
\caption{The detector, consisting of four 1 $\times$ 1 m X-Y (scintillator bar) Units and a toroidal magnet. 
The DAQ box looks for a coincidence in the X-Y Unit and sends the data to the PC. 
The clocks of the four DAQ boxes were synchronized by a clock generator module.
}

\label{fig:system}
\end{figure}

The detector system illustrated in Fig. \ref{fig:system} 
consisted of four X-Y Units with a 61-cm long 1.4-T permanent magnet located between the second and third X-Y Units. 
Each X-Y plane was constructed from two detector planes arranged at 90$^\circ$, each plane consisted of 100 scintillating strips of width 1~cm.
The distance between Unit-1 and Unit-2 (composing the Front Tracker) was 1.5 m, 
whilst the distance between Unit-3 and Unit-4 was 0.75 m (compromising between the acceptance of cosmic muons and the momentum resolution). 
The elevations of the four X-Y Units were adjustable with respect to the toroidal magnet. 

\begin{figure}[htbp]
\begin{center}
\includegraphics[width=11.0cm, bb=0 0 720 540]{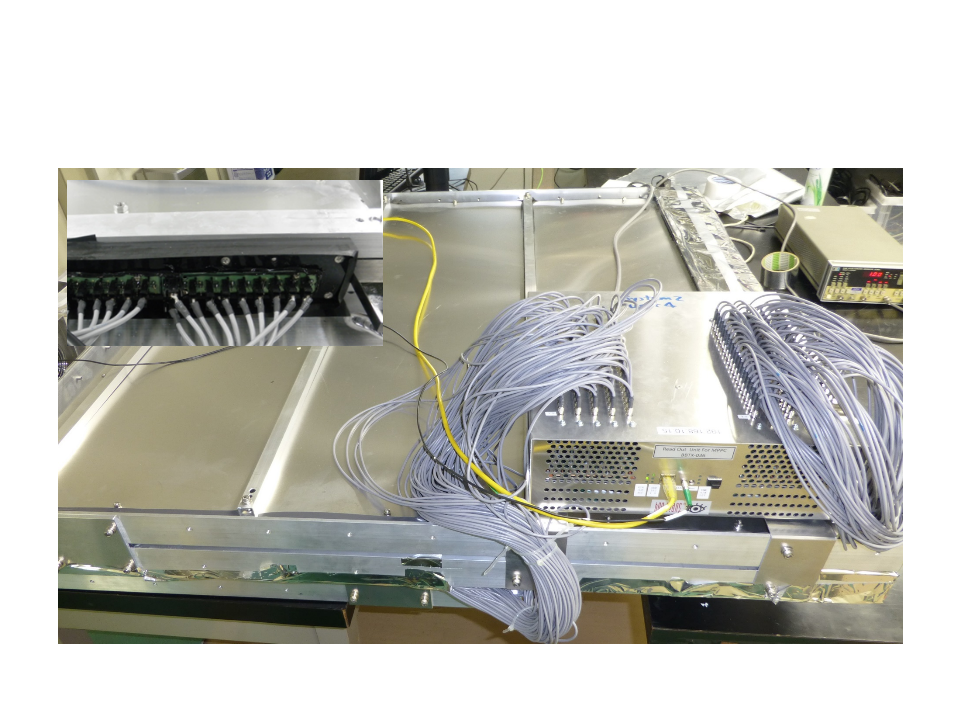}
\end{center}
\caption{Photograph of one XY unit. The DAQ box collects 200 signals from X and Y planes.
The inset is an enlarged view of printed boards each serving connection of 10 MPPCs to the DAQ box via 10 coaxial cables.}
\label{fig:XYunit}
\end{figure}

Each X-Y Unit (see Fig.~\ref{fig:XYunit}) was equipped with a data-acquisition (DAQ) box that collected 200 MPPC signals; 100 each of X and Y. 
The X-Y coincidences (after applying thresholds to the signals in a given time window) were transferred via a LAN to a PC 
where the trigger decision was made. 
The timing of the four DAQ boxes was reset every 1~$\mu$s using a common clock generator that provided
NIM-logic signals to all four of the DAQ boxes. 
The whole system was housed in a 20-ft shipping container, 
reinforced with steel bars, equipped with air conditioning, and which weighed 3.5 t.  
The total weight of the system amounted to 10~t including the toroidal magnet (5.3 t) and four X-Y Units (each 50 kg).

\subsection{The X-Y scintillator units}

The eight detector planes, each consisting of 100 scintillating bars, were identical in design. 
The base element consisted of scintillator\cite{elgin} bars of width 9.7 mm, height 6 mm (nominal), and length 1000 mm. 
We machined a 2 mm deep straight groove where a 1 mm $\phi$ wavelength shifting fiber 
(Kuraray Y11 200 ppm doped double clad plastic fiber\cite{kuraray}) was embedded. 
After cutting to a length of 1010 mm, both ends of the fiber were polished and one end was aluminized. 
The aluminum mirror was reinforced with white paint\cite{paint}, 
and the three sides of the bar were also painted white. 
Twenty such bars were placed with the grooved faces set against a jig plate 
(where 20 ridges were machined at a spacing of 10 mm), such that the bars were arranged at a spacing of precisely 10 mm. 
Five sets of such 20-bar arrangements were put in a light-tight tray, made of 8 mm thick, black PMMA bars at three sides, 
with a 0.5 mm thick aluminum sheet screwed to the PMMA bars. 
A 0.02 mm thick white PET sheet and a 0.1 mm thick black vinyl sheet were used for light reflection and light shielding, respectively. 
In order to keep the shape and flatness of the tray, four 20 $\times$ 30 mm rectangular aluminum tubes were screwed 
in a direction perpendicular to the scintillator bars (see Fig.~\ref{fig:XYunit}). 

For the fiber end-plate, one hundred holes were machined at a spacing of 10~mm to a 10 mm thick black PMMA end-plate, 
where wave-shifting fibers were inserted and glued. 
After placing the fibers into the scintillator grooves and setting the white PET and vinyl sheets, 
a 2 mm thick foam sheet was inserted, after which the tray was covered with another 0.5 mm thick aluminum sheet. 
The foam was inserted for the purpose of absorbing any thickness variation of the scintillator bars. 
The open side of the tray box was enclosed by screwing the fiber end-plate against a pair of enforcement bars. 
The end-plates were machined with screw holes and five dowel pins, 
to which an MPPC array plate was then positioned and screwed.

\subsection{The MPPC array plate}

\begin{figure}[htbp]
\begin{center}
\includegraphics[width=15.0cm, bb=0 0 825 400]{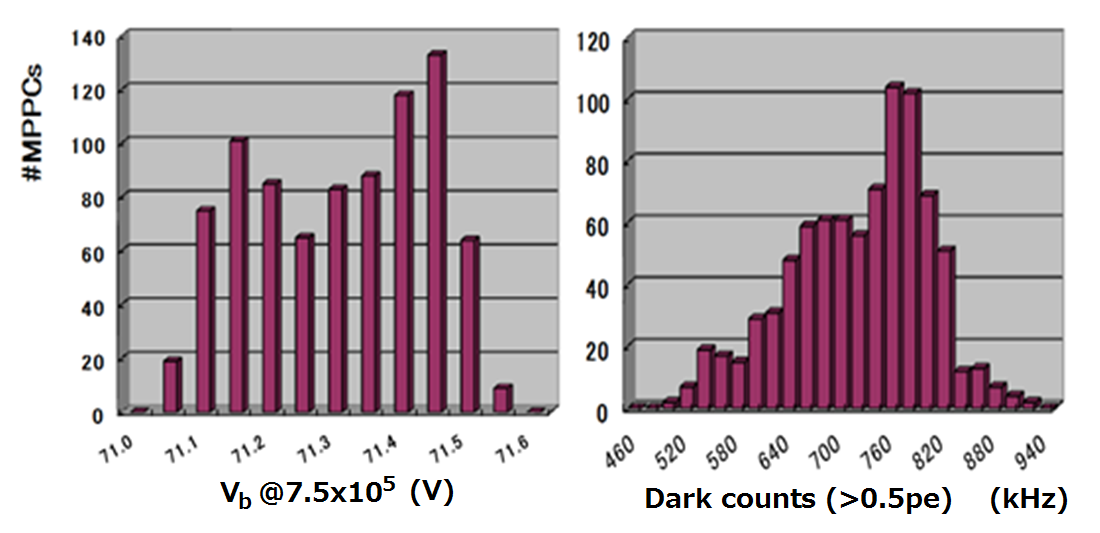}
\end{center}

\caption{(Left) Distribution of bias voltages, V$_{\rm b}$ at a gain of 7.5 $\times$ 10$^5$, and (Right) dark count rates above a threshold of 0.5 PEs. 
Data were taken at 25$^\circ$C for all 840 delivered pieces.}
\label{fig:mppc}
\end{figure}

\begin{figure}[htbp]
\begin{center}
\includegraphics[width=5.0cm, bb=0 0 640 1097]{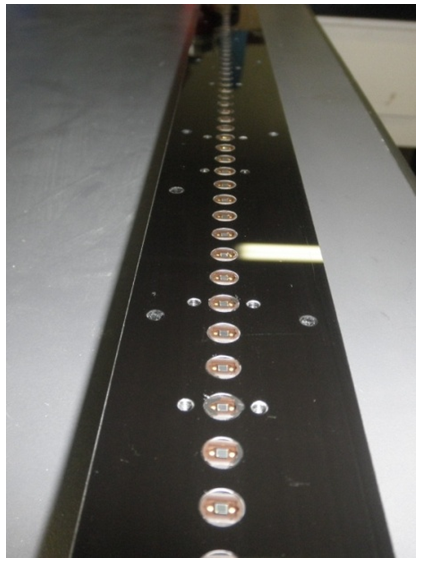}
\end{center}
\caption{Photo-coupling side of the MPPC array plate. The MPPC ceramics were fit into holes of 6 mm diameter. 
The innermost rectangles are MPPC sensors of 1.3~mm square. }
\label{fig:mppcpl}
\end{figure}

The multi-pixel photon counter (MPPC) is a product of Hamamatsu Photonics\cite{hpk} that 
has 667 pixels covering a 1.3 mm square area and a typical gain of 7.5 $\times$10$^5$ at 71 V bias. 
The same devices were developed for the T2K near detector\cite{t2k} located at KEK. 
Fig.~\ref{fig:mppc} shows the bias voltage distribution at a gain of 7.5 $\times$10$^5$, and the dark count rates 
for a 0.5 photoelectron (PE) threshold for all of the 840 MPPCs used here. 
The devices showed quite uniform characteristics. 
The dark count rates decreased to 38-86 kHz at a threshold of 1.2~PEs.
The MPPCs were packaged in 6 mm $\phi$ ceramics with the photo-coupling face covered by a 0.3 mm thick silicon resin. 
Each of the wave-shifting fibers was coupled to one MPPC 
so that 100 MPPCs were aligned at spacing of 10 mm along the 5 mm thick black PMMA plate.  
The MPPC array plate was assembled by first placing each MPPC in drilled holes (see Fig~\ref{fig:mppcpl}). 
The MPPC anode and cathode leads were then soldered onto a printed circuit board. 
One circuit board, serving for 10 MPPCs, was mounted with 10 2-pin headers (Hirose Electric Co., Ltd., HNC Series),
see the inset of Fig~\ref{fig:XYunit}. 
The signal was transferred via coaxial cables coupled individually to the two leads of the MPPC, 
hence the signal pulse was overlaid on a typical bias voltage of 71~V.

\subsection{The MPPCs DAQ system}

\begin{figure}[htbp]
\begin{center}
\includegraphics[width=15.0cm, bb=0 0 771 546]{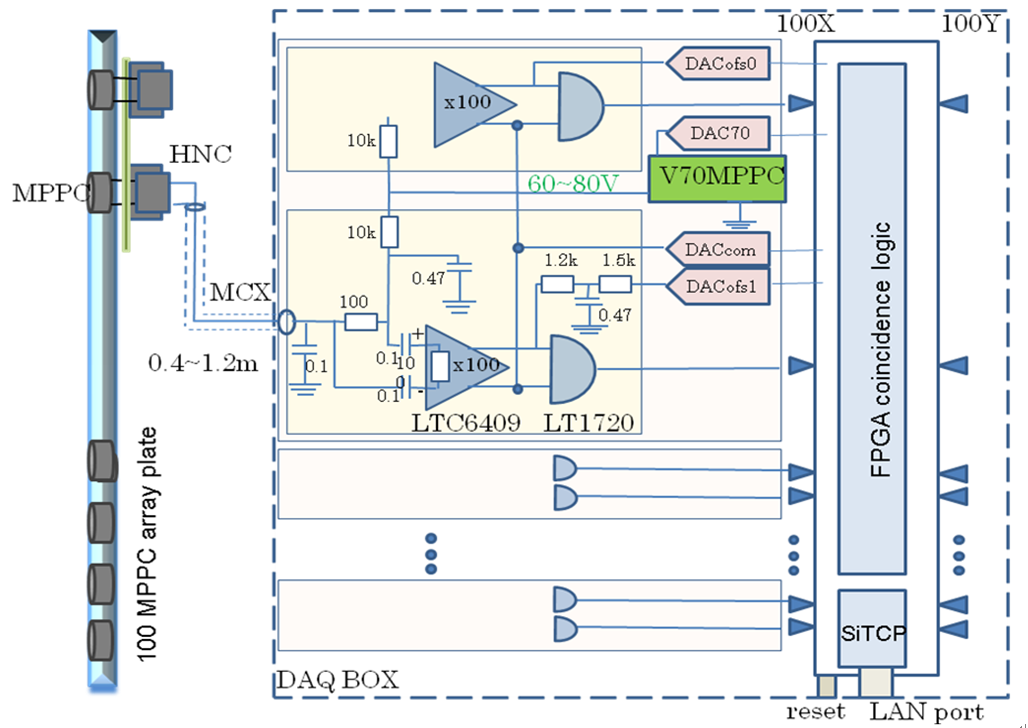}
\end{center}
\caption{Schematics of the DAQ boxes that handle 100 X and 100 Y analog signals. 
The bias voltage, controllable via DAC70, is common to two neighboring channels whilst the threshold is adjustable via DACofs0 and DACofs1 individually. The DAC settings and data readouts were performed through a single LAN port.
}
\label{fig:daq}
\end{figure}

The MPPCs data acquisition (DAQ) system was fabricated by Bee-Beans Technology\cite{bbt1}. 
Each system dealt with 200 analog signals for the two planes of one X-Y Unit, and also provided bias voltages to the MPPCs themselves. 
A block diagram of this system is shown in Fig.~\ref{fig:daq}. 
After decoupling the DC bias voltage, the pulse signal was transferred to an LTC6409 fast differential amplifier with its gain set to 100. 
The amplified signal was then discriminated using an LT1720 functioning in zero-crossing mode, 
where the baseline (i.e., threshold adjustment) is provided by an individual digital-to-analog converter (DAC). 
Twenty amplifier and discriminator stages were grouped, and assembled on a sub-board. 
Ten such sub-boards were plugged into a motherboard furnished with FPGAs (for coincidence logic generation and SiTCP communication). 
The motherboard also controls UDP commands to the DACs via the LAN port. 
The system ran on a 125~MHz clock cycle, providing a shortest coincidence window of 8~ns. 
The FPGA looks for a single cluster in both the X and Y planes in a given time window (adjustable in units of 8 ns up to 1~$\mu$s). 
This cluster is defined as hits in consecutive channels from `one' up to `one to four' (selectable) and
we chose `one to three' consecutive hits in order to allow for inclined tracks.
 Once this coincidence condition was met, the hit positions (in X and Y) and the timestamps were buffered whilst waiting to be transferred. 
The timestamp clock was reset and the four DAQ boxes in use were synchronized by external pulses, provided by
the clock distributor\cite{bbt2} which was operated at 1~kHz.

The signal line lengths in the DAQ box were designed to be identical so that the propagation inside did not bias the coincidence judgment. 
The propagation time difference from the scintillator to the DAQ box input was 10~ns at maximum and 
we verified that a coincidence window of 16~ns was appropriate for the X-Y Units. 
The Unit coincidence data (X and Y hit positions, and timestamps) were transferred to a PC, 
where they were examined for coincidence in Unit-1 and Unit-2 over a 32~ns time window. 
If the event was judged acceptable, these Unit coincidence data were stored together with 
those from Unit-3 and Unit-4 that had the closest timestamps.

\subsection{Response uniformity of the detector}\label{nonuniform}

The light yield of the scintillator/fiber-MPPC system was typically 15 photoelectrons for $\beta$ rays 
penetrating near the scintillator end opposite to the readout end, which is comfortably high to detect muons.
The threshold voltages of all 800 channels were at first adjusted to discriminate the noise contribution. 
The response uniformity in a plane was then examined using muons recorded as coincidence in the X-Y Unit placed horizontally.
We accumulated about 10k events per channel where events with only one hit in each plane were accepted.
The rms variation of the counting uniformity ranged from 16\% to 19\%.
The obtained non-uniformity includes the effects of cross talk to neighboring scintillators.

\subsection{Momentum measurement}

Measurement of the muon momentum provides additional information on the remote mass distribution and 
should help improve the radiography image by choosing tracks in an appropriate momentum range. 
Also, the muon momentum information was indispensable in simulations to predict the obtained image. 
In light of this we constructed a permanent toroidal magnet with a magnetic field of 1.4~T 
in collaboration with NEOMAX Engineering\cite{neomax}. 
As the results using the measured momentum information are beyond the context of this paper and are 
to be reported in a separate paper; we will not discuss this issue further here.

\section{Test Experiment at the JAPC}

\subsection{Detector installation and operation}
 
The system was transported to the JAPC on February 24$^{\rm th}$, 2012 and installed outside the reactor building at a 
distance to the reactor center of 64 m at an angle of 22$^{\circ}$ with respect to one of the building's walls. 
Although the detector started data acquisition on the same day, some detector threshold adjustments were carried out for the first few days. 
In the period up to June 18$^{\rm th}$, 2012 (referred to here as PERIOD-1), the detector system was aimed at the center of the fuel loading zone 
by adjusting the relative heights of Unit-1 and Unit-2. 
In the period from June 28$^{\rm th}$ to October 5$^{\rm th}$ (referred to here as PERIOD-2), the elevation of Unit-1 was raised, thus
aiming the detector at the fuel storage pool.

\subsection{The coincidence time distribution} 
\begin{figure}[htbp]
\begin{center}
\includegraphics[width=8.0cm]{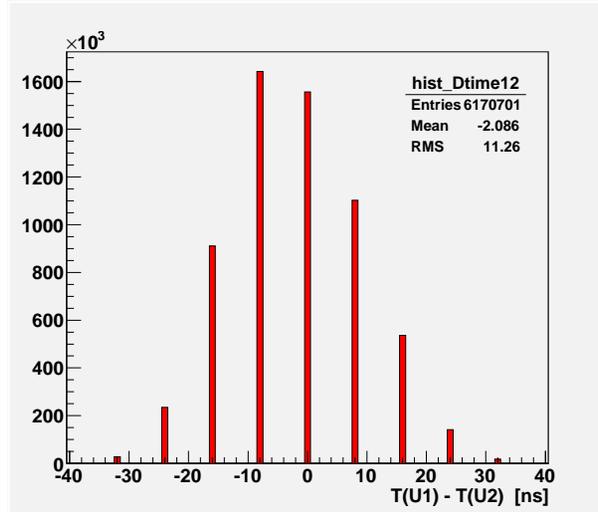}
\end{center}
\caption{Time difference between Unit-1 and Unit-2 in PERIOD-1.  The time was digitized in units of 8 ns.}
\label{fig:dtime}
\end{figure}

The time window to judge an X-Y coincidence inside the same Unit was fixed to 16 ns. 
The time difference between Unit-1 and Unit-2 was then used to judge whether the event was to be recorded or not. 
As seen in Fig.~\ref{fig:dtime} where the time difference,
 T(Unit-1)$-$T(Unit-2), is plotted, 
the mean of the time difference distribution was shifted from zero,
reflecting that tracks coming from the front were dominant in this configuration. 
This asymmetry was caused by the different zenith angle coverage for the tracks coming from the front and back 
(Unit-1 was elevated by 20 cm in PERIOD-1), and by the magnet behind.
Events within a  32~ns time window were retained.

\subsection{Coincidence rates}

\begin{figure}[htbp]
\begin{center}
\includegraphics[width=7.0cm, bb=0 0 380 298]{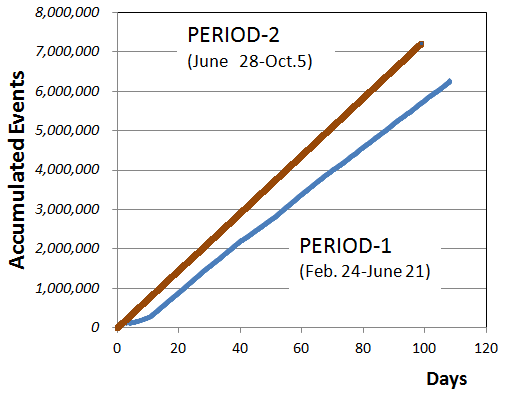}
\includegraphics[width=7.0cm, bb=0 0 401 318]{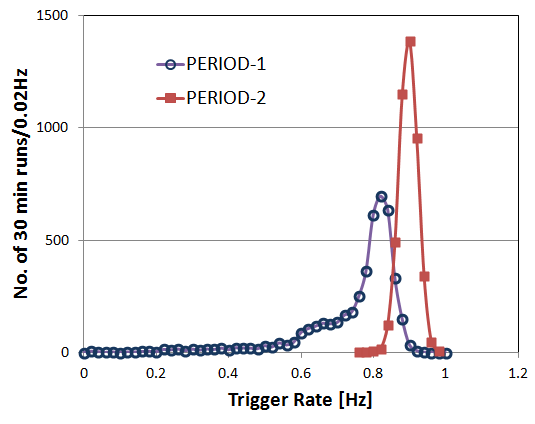}
\end{center}
\caption{(Left) Number of recorded events versus the number of measurement days, and (Right) trigger rate calculated every 30 min. 
The data were accumulated in two periods; PERIOD-1 and PERIOD-2.}
\label{fig:accum}
\end{figure}

Fig.~\ref{fig:accum} summarizes the number of recorded events and the trigger rates during PERIOD-1 and PERIOD-2.
In total, 6.3 million events and 7.1~million events were accumulated in 
PERIOD-1 (over 114 days) and PERIOD-2 (over 99 days), respectively.
In both periods the accumulated number of events increased linearly with days (except for initial threshold adjustments period in PERIOD-1), verifying that the detector was operated stably in the period over 7 months. 
Each data acquisition run lasted 30 min and the trigger rate in Fig.~\ref{fig:accum} is calculated for each run.
In PERIOD-1, the trigger rate showed a peak at 0.85 Hz, and a tail in the lower frequencies. 

The MPPC gain is temperature dependent and decreases typically by 2.2\% per degree. 
The tail was a result of this gain variation against fixed threshold, 
which happened to be unexpectedly wide due to a large variation 
of the room temperature from 10$^\circ$C to 28$^\circ$C caused by a lack of thermal insulation against the container walls. 
Towards the end of PERIOD-1, thermal insulation of the container was accomplished.
Also at the beginning of each data acquisition run, the room temperature was used to scale the threshold values for the MPPCs.
As evident, the trigger rate was very stable in PERIOD-2, averaging at 0.9~Hz. 

The reduced trigger rate in PERIOD-1 was associated with non-uniform detector efficiency additionally to the
response non-uniformity described in \ref{nonuniform}. 
Such non-uniformity, though, has little impact on the image of the reactor 64~m away.
This is understandable since the track projections with the same differences of hit scintillator numbers in the two tracking units are localized within 1~m, therefore the reactor image where we employed a bin size of 40~cm is obtained essentially integrating over the local non-uniformity of the detector response.
The above estimate was verified by examining the projected image using tracks coming from behind; 
no structure other than the acceptance was created since no massive structure was present.
We also have verified that these low trigger rate runs did not introduce any spurious projections of tracks. 
After these studies, we only selected runs with a trigger rate above 0.5 Hz for further analysis.

\subsection{Track distributions}
 
\begin{figure}[htbp]
\begin{center}
\includegraphics[width=9.0cm, bb=15 0 579 305]{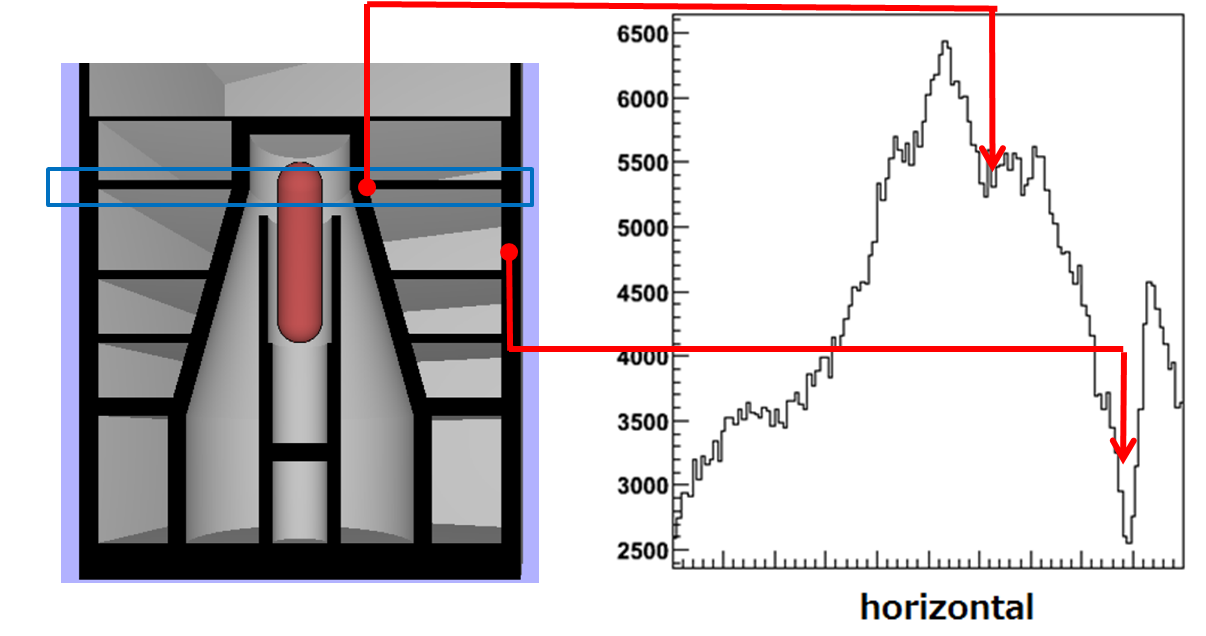}
\includegraphics[width=9.0cm, bb=15  0 558 303]{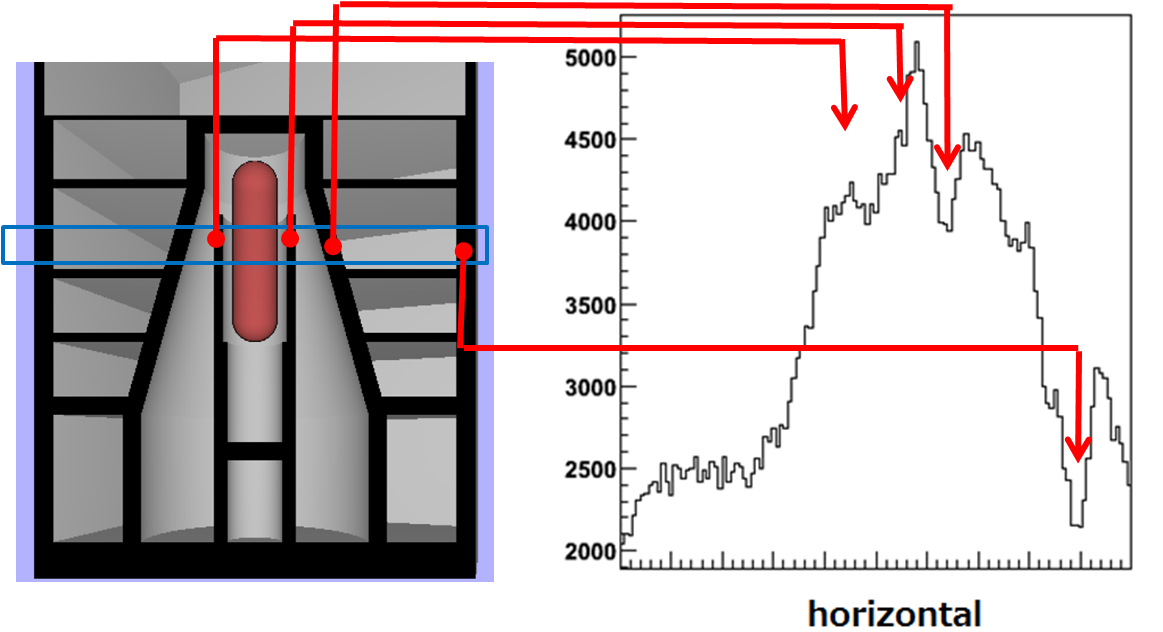}
\includegraphics[width=9.0cm, bb=15 0 579 305]{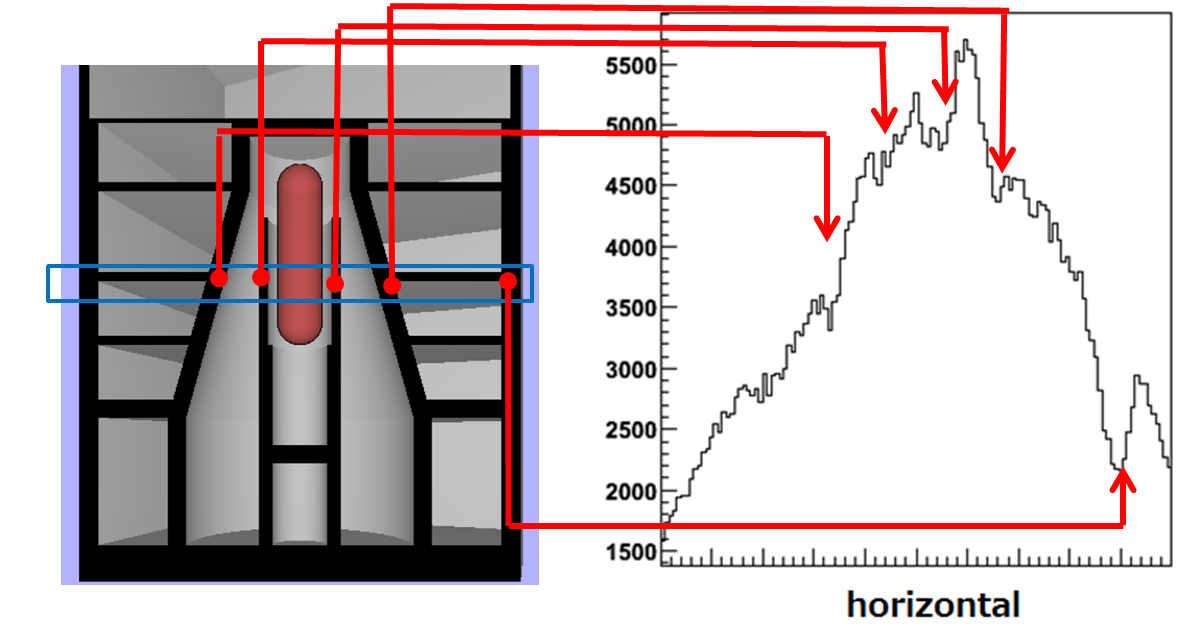}
\includegraphics[width=9.0cm, bb=15 0 579 305]{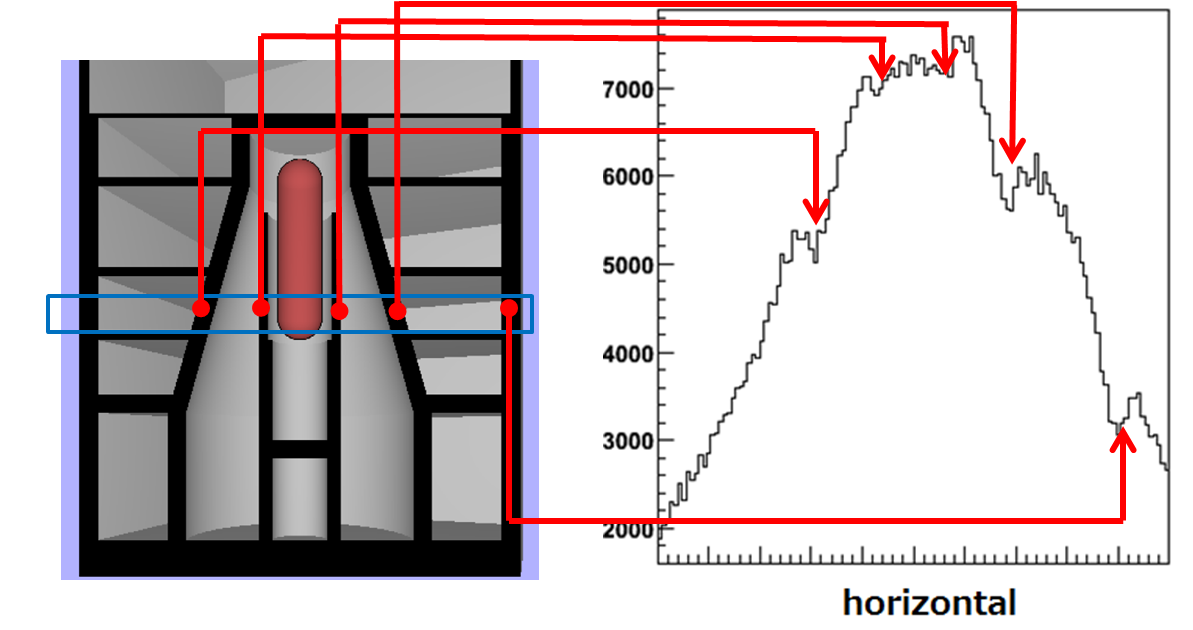}
\end{center}
\caption{Horizontal track distributions projected at the reactor center for 4~m vertical slices.
The vertical regions cover the areas (from bottom to top) just below the fuel loading zone, the fuel loading zone, just above the fuel loading zone, and further above.  
The recognizable dips correspond to the right-side building wall, containment vessel walls, and pressure vessel support walls. 
Because of the angle of the detector with respect to the reactor building, the projection of the left-side wall was wider in $X$, 
hence no corresponding dip was recognizable. 
}
\label{fig:height}
\end{figure}

Since the DAQ system selected only events with a single cluster of hits per plane, 
tracking was straightforward. 
The reconstructed tracks obtained 64~m away from the reactor center 
were extrapolated to the reactor center plane (defined as $Z=0$).
The plots in Fig.~\ref{fig:height} 
show the track distributions along the $x$-axis for given slices in the $y$-axis at $Z=0$, alongside 
the corresponding reactor structures available in public \cite{plant}.

\subsection{Two-dimensional image}

\begin{figure}[htbp]
\begin{center}
\includegraphics[width=8.0cm, bb=0 0 500 465]{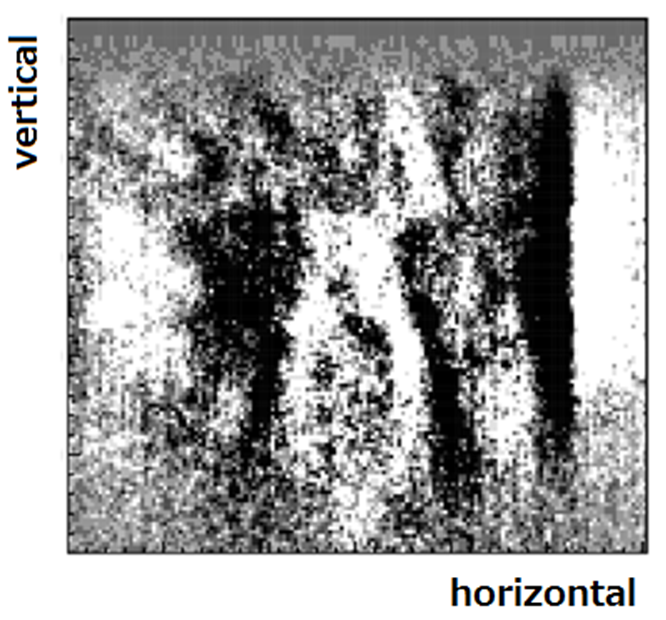}
\end{center}
\caption{Image obtained during PERIOD-1. The massive structures are represented by darker parts.}
\label{fig:fourier}
\end{figure}

Two-dimensional images of detected muon tracks extrapolated to the reactor center are 
illustrative to demonstrate the ability of the muon radiography.
As obtained images, though, are affected by the massive floors that prevent imaging of detailed structure of the fuel region we are interested in.
We present a two-dimensional image as shown in Fig.~\ref{fig:fourier} where 
the visibility of the image was improved by subtracting such slowly changing components. 
In practice, we expanded the horizontal distributions of each vertical bin 
to a Fourier series 
and ignored the low frequency components, 0$^{\rm th}$ to 3$^{\rm rd}$ series of the expansion, on reconstruction of the image.
Through this procedure the effect of the geometrical acceptance is also suppressed.

In the image, the darker parts correspond to massive structures. 
The conical-shaped containment vessel was visible together with the cylindrical reactor pressure vessel (RPV) at its center.
Also, the right side building wall was clearly seen, as it was almost projective to the detector. 

\subsection{Sensitivity evaluation using the data}

Evaluation of the detector sensitivity using only the available data is described in this section.
We present the ability to detect variation in the pool water level and the determination of the spatial resolution of the detector system. 

\subsubsection{Detection of water level variation}\label{WLvariation}
 
\begin{figure}[htbp]
\begin{center}
\includegraphics[width=8.0cm]{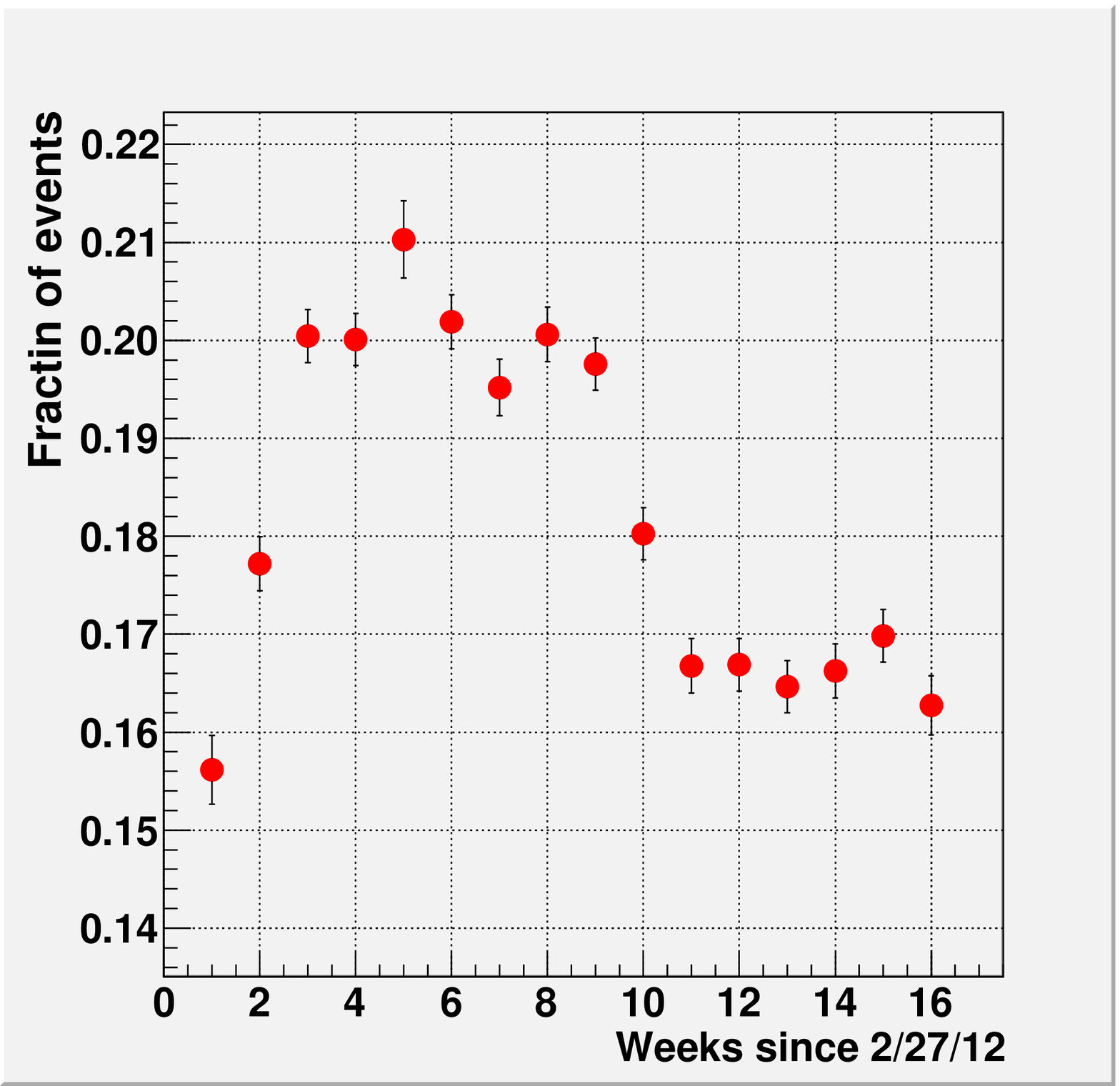}
\end{center}
\caption{Event yield in the 4 $\times$ 9~m region corresponding to the dryer-separator pool, divided by the event yield in the same 4~m horizontal band.
The $x$-axis shows the weeks passed since Feb 27$^{\rm th}$, 2012.
}
\label{fig:weekbump}
\end{figure}

To investigate the possible variation of the reactor plant configuration, we first examined the distribution uniformity on a weekly basis, and 
found a distinct variation of the event yields in specific regions. 
Fig.~\ref{fig:weekbump} plots the event yield in a 4~m (horizontal) by 9~m (vertical) region, corresponding to the dryer-separator pool,
divided by that in the same 4~m horizontal band, shown in a weekly basis. 
The event rate was systematically high from Week 3 to Week 8, with transitions seen at Week 2 and Week 10. 
According to the work record provided after this study was made, 
the water level in the reactor well had changed during this period: 
reducing to $-7$~m on March 9$^{\rm th}$ (Week 2), increasing up to $-5$~m on April 27$^{\rm th}$ (Week 9) 
and $-2.6$~m on May 1$^{\rm st}$ (Week 10), then back to the nominal 0~m on May 6$^{\rm th}$ (Week 10). 
This variation in the water level (WL) explains the transitions seen in Fig.~\ref{fig:weekbump}.

\begin{figure}[htbp]
\begin{center}
\includegraphics[width=10.0cm, bb=0 0 636 593]{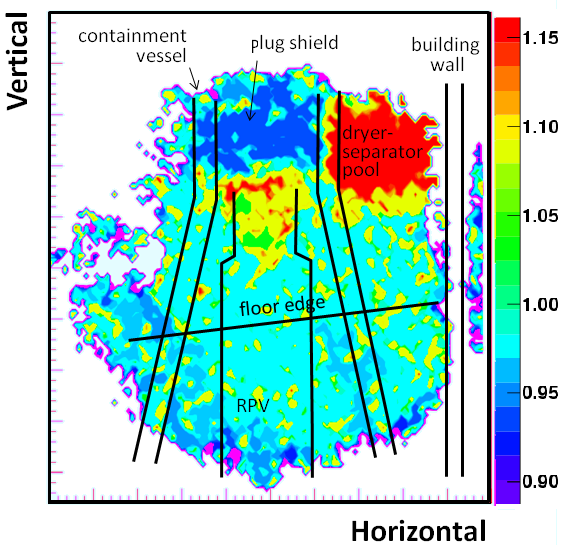}
\end{center}
\caption{
Normalized ratio of the event yields whilst the reactor plug shield was `in' and WL$=-7$~m (at 31 days), 
to the event yields whilst the plug shield was `out' and WL$=$0~m (at 45 days). 
Some of the plant structures are overlaid in black, and data points with poor statistics are excluded.
A full description of the color code is given online. 
}
\label{fig:plugON}
\end{figure}

We were also informed by the JAPC that the well shield plug was `in' between 
March 20$^{\rm th}$ and April 21$^{\rm st}$ and `out' for the other periods. 
Whilst the plug was `in', the WL reduced to $-7$~m. 
Fig.~\ref{fig:plugON} shows a two-dimensional ratio of the event yield between the `plug-in' period and 
the `nominal' period (plug was out, and WL$=$0~m). 
The upper-right part of the figure corresponds to the dryer-separator pool.
The most distinct variation in the figure was due to a change in the water level of the dryer-separator pool. 
The low ratio part at the top of containment vessel was due to the plug shield (1.7 m thick concrete), 
and the higher ratio part below the shield was due to the reduced WL in the reactor well. 
Since the detector was located at an angle to the building,
the same heights of the WL in the dryer-separator pool and the reactor well did not manifest as horizontal features in this elevation view, 
as can be seen from the overlaid floor structure in Fig.~\ref{fig:plugON} 
and an density-length distribution plot given in Sec.~4 (Fig.~\ref{fig:densityLength}).

\begin{figure}[htbp]
\begin{center}
\includegraphics[width=8.0cm]{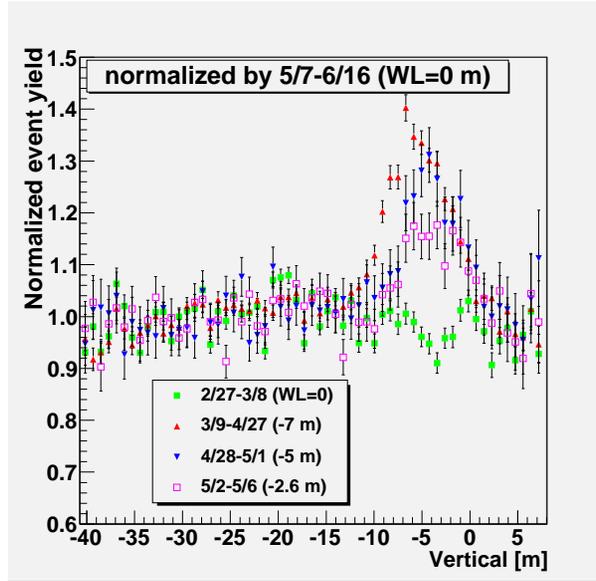}
\end{center}
\caption{Vertical yield distributions for periods with different water levels, 
normalized by the distribution at the nominal water level (WL$=$0, where the vertical origin is set). 
The horizontal range covers the dryer-separator pool.
}
\label{fig:WL}
\end{figure}

Fig.~\ref{fig:WL} shows the vertical yield distributions for the periods with different WLs of the dryer-separator pool, 
normalized by the distribution obtained when WL$=$0~m (May 5$^{\rm th}$ to June 16$^{\rm th}$).
The data points in the horizontal range corresponding to the pool region were added. 
The widths of the bumps reflect the reduced water level. 
Note that the data period was only 5 days for two of the four periods. 
The dryer-separator pool was 12~m thick along the muon path, which is equivalent to a 4.8~m thick fuel material 
(for a density of 2.5 g/cc). 
The observation of the WL changes can be interpreted as the detection ability of the corresponding thick fuel
material in the corresponding period.

\subsubsection{Spatial resolution}

\begin{figure}[htbp]
\begin{center}
\includegraphics[width=8.0cm, bb=0 0 1039 815]{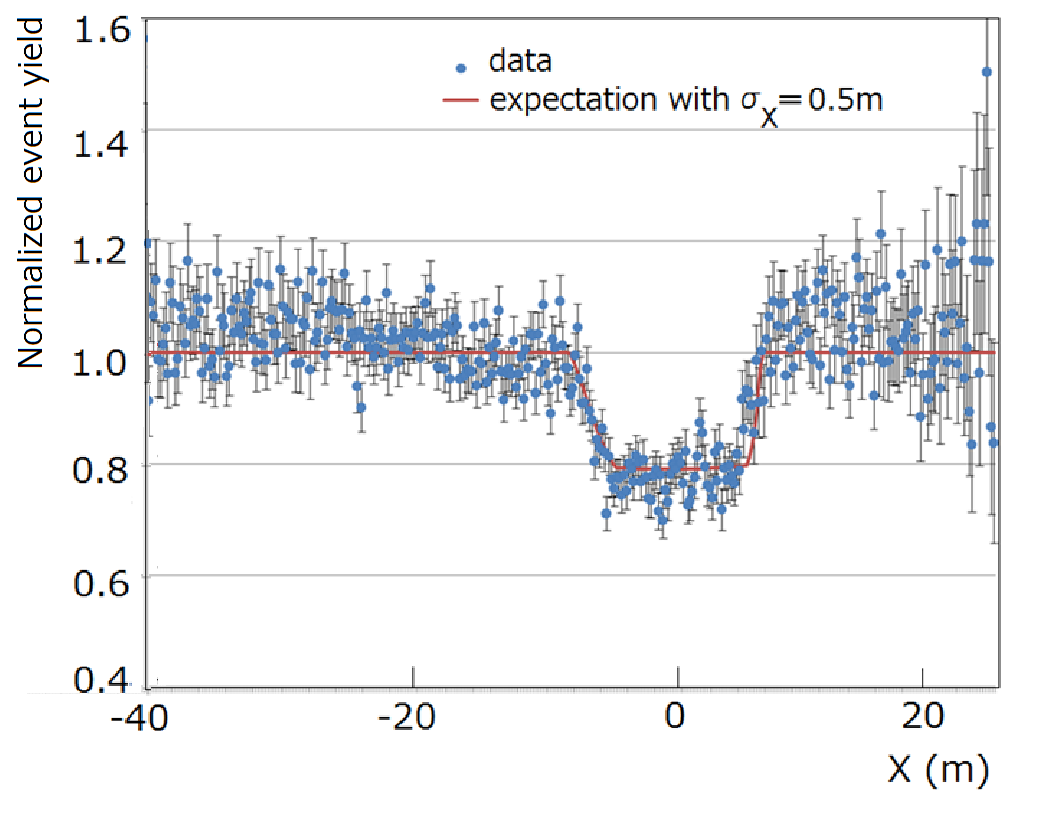}
\end{center}
\caption{Horizontal event yield distributions at the dryer-separator pool height, 
where the distribution at normal water level (WL$=$0~m) was normalized by that at a reduced water level (WL $<$0).
The horizontal origin is set at the center of the pool. 
The overlaid curve indicates the expected results where the spatial resolution was assumed to be 0.5 m.
}
\label{fig:sigmaXpool}
\end{figure}

The spatial resolution is an important parameter if we are to locate the fuel material. 
Here we use the data distributions to evaluate the spatial resolution. 
The horizontal yield distribution across the dryer-separator pool is plotted in Fig.~\ref{fig:sigmaXpool}, 
where the distribution at WL$=$0~m was divided by that at a reduced water level (WL $<$0); 
the dip observed here is due to the presence of water.
 The dip distribution was compared to the expected distribution where the spatial resolution was assumed to be 0.5 m, and showed good  
agreement, certifying the validity of this assumption for the spatial resolution.

\begin{figure}[htbp]
\begin{center}
\includegraphics[width=12.0cm]{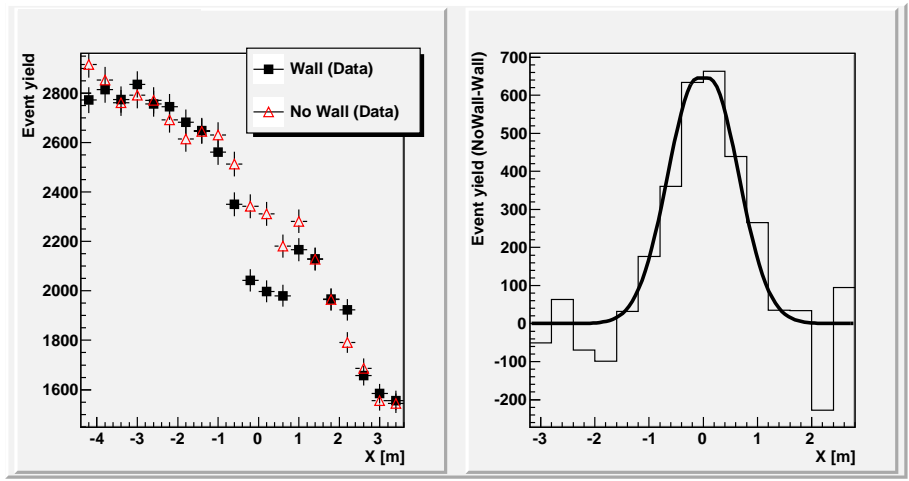}
\end{center}
\caption{(Left) Horizontal event yield distribution at a vertical range including the wall under investigation (black) 
in comparison to that below where the wall was not included (red triangles). 
(Right) Flat-topped Gaussian fit of the difference between the two curves.
}
\label{fig:sigmaXwall}
\end{figure}

Another evaluation of the spatial resolution was obtained by looking at the width of a wall. 
One of the walls supporting the dryer-separator pool was 1.25 m wide and was viewed as an almost projective plane by the detector. 
The event yield distributions in the vertical bands both with, and without the wall, are shown in Fig.~\ref{fig:sigmaXwall}. 
By taking the difference of the two distributions and applying a Gaussian fit, 
we obtained a spatial resolution of 56 cm with a wider wall width 1.31 m. 
Since the wall was 13~m thick along the muon path and thus not completely projective, 
the difference obtained from this data was then fitted to a flat top Gaussian, as shown in Fig.~\ref{fig:sigmaXwall}. 
This fit resulted in a spacial resolution of 49.1 $\pm$ 2.7~cm.

\section{Comparison of the Experimental Results with Simulations}

\subsection{The density-length method}\label{dlm}

A simple method to evaluate the material along the muon path was developed using the density-length method 
(the product of the density and path length). 
The assumptions made were;
\begin{enumerate}
\item
The observed muon flight direction was identical to the direction of the original muon entering the structure (no scattering).
\item
The muon absorption was determined from the sum of the density-lengths of the materials along the path. 
(The absorption was an exponential function of the density-length sum.)
\item
The original muon intensity was dependent only on the zenith $\theta$, and azimuth $\phi$ angles. 
The intensity was uniform across the azimuth angles, whereas  
the zenith angular dependence of the intensity was assumed to follow $\cos^2\theta$.
\item There was a uniformly distributed background which was independent of the zenith angle.
\end{enumerate}

We incorporated these assumptions in the following equation:
\vspace*{-5mm}
\begin{center}
\begin{equation}\notag
 {\it N}_{\rm obs}(\theta, \phi)=N_\mu(\theta, \phi) \cdot \exp[-\sum \rho_i t_i(\theta, \phi)/\lambda ] + n_{\rm bk}
\end{equation}
\end{center}

\noindent
where  $ N_{\rm obs}(\theta, \phi)$ is the observed muon intensity in the direction $(\theta, \phi)$, 
$N_\mu$ is the original muon flux entering the structure, 
$\lambda$ is the effective absorption length, 
$\rho_i$ is the density of the $i$-th material, 
and $t_i(\theta, \phi)$ is the length of the $i$-th material along the muon path. 
In practice, the angles were defined by the horizontal and vertical coordinate differences $(l,m)$, 
and their acceptances were not uniform in $(\theta, \phi)$ 
due to geometrical effects and muon flux. 
The acceptance functions $f(l,m)$ and $g(l,m)$ in terms of the direction bin 
were calculated by a Monte-Carlo method 
so that the above relation was reduced to: 

\begin{center}
\begin{equation}\notag
 {\it N}_{\rm obs}(l,m)= a \cdot f(l,m) \cdot \exp[-t(l,m)/\lambda] + b \cdot g(l,m)
\end{equation}
\end{center}

\noindent
where  $N_{\rm obs}$ is the number of muons observed in the direction bin $(l,m)$,
$f(l,m)$  is the $\cos^2\theta$ dependent acceptance function, and 
$g(l,m)$ accounts for pure geometrical acceptance of the background.

\begin{figure}[htbp]
\begin{center}
\includegraphics[width=6.5cm, bb=0 0 703 605]{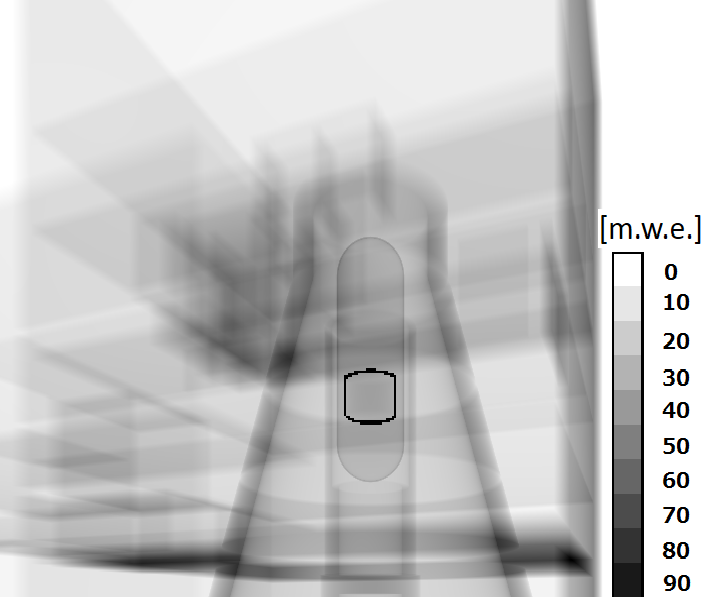}
\includegraphics[width=8.2cm, bb=0 0 519 301]{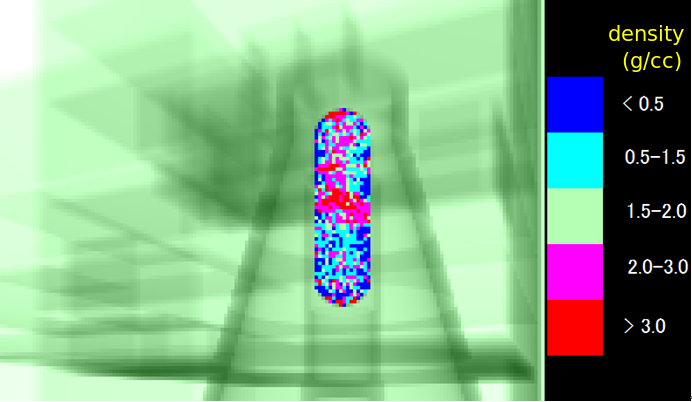}
\end{center}
\caption{(Left) Simulated density-length distribution of the reactor. 
The scale is in m.w.e. (meters water equivalent). 
The outline of the fuel is highlighted.
(Right) Evaluated density distribution inside the RPV. 
The density was consistent with 1 g/cc over the entire fuel loading zone. 
}
\label{fig:densityLength}
\end{figure}

Fig.~\ref{fig:densityLength} shows a two-dimensional density-length distribution calculated by this method. 
The fuel was assumed to be loaded in the RPV.
In addition to the containment vessel and RPV, some vertical walls and horizontal floors are recognizable.

The density distribution inside the RPV was determined by comparison with the observed event yield, 
while fixing the density-lengths of the other structure components relying on the input values.   
In this calculation, the ratio of the coefficients $a$ and $b$ were fixed, leaving only the normalization as a free parameter.
Since the variation in absorption can be approximated as a linear function for small density variations, 
the event yields for common RPV densities of 0 and 3 g/cc were calculated in each of 40 $\times$ 40 cm mesh. 
These data-sets were used to interpolate to estimate the density in that mesh (which is uniform along the muon path) 
by comparison to the observed yield. 
The evaluated density distribution is shown in Fig.~\ref{fig:densityLength}, 
which resulted that the fuel loading zone was consistent with a density of 1 g/cc. 
There were some higher density parts in the upper half of the RPV that we
anticipate were due to contributions (such as scattering at the floor and the wall) that were not completely evaluated.

\subsection{Monte Carlo methods}
 
The GEANT simulation\cite{geant4} was employed to simulate the muon interactions in the reactor and the building structure. 
Since the available plant information was limited, the building walls and floors; containment vessel and RPV; and the pools 
were the main components implemented in the simulation. 
The thicknesses and locations were taken from a drawing \cite{plant} 
where it was assumed that they were concrete with a density of 2.5 g/cc, except for the RPV wall and the water. 
The RPV was assumed to be either empty, filled entirely with water, 
or filled with water and with fuel material loaded.

The GEANT simulation traces a muon of given momentum and direction through a given material. 
Therefore, every critical material component needs to be taken into consideration, 
and reliable muon momentum and direction distributions are indispensable. 
The momentum distribution was taken from Jokisch et al.\cite{jokisch} who measured the distribution at a 75$^{\circ}$ zenith angle. 
For the zenith angular dependence of the muon yield, 
we employed the data by the Okayama group\cite{okayama} who measured the cosmic muon flux at sea level for the zenith angular range from 0 to 80$^{\circ}$. 
They provided the zenith angular dependencies for three muon momentum bins of 12.5, 22.5, and 45 GeV/c. 
Their zenith angular dependencies were fitted to a $\cos^n\theta$ function, 
the resultant $n$ determined as 1.09, 0.76, and 0.50, respectively. 
These numbers were used to obtain the momentum ($P$) dependence of $n(P)$ as:

\begin{equation}\notag
n(P) = 3.63 \cdot (P/{\rm (GeV/c)})^{-0.505}
\end{equation}

Processing of GEANT events is time consuming as it requires an optimization of event production.  
We carried out a study to generate events randomly at $Z=0$ that then were traced from a position behind the reactor building. 
The scattering angle distribution for the events that were accepted by the detector was found to be within 0.03 rad. 
To summarize, we first picked up the momentum of the muon according to Jokisch {\it et al.}, 
then generated events randomly at $Z=0$ with the zenith and azimuth angles given in the above described way. 
This generation was repeated until the track angle with respect to the detector center was within 0.03 rad. 
Successful events were traced through the building and reactor material. 
This event production method substantially improved the reproduction of the reconstructed track distributions, 
compared to a method where only the tracks that would have passed through the detector were traced 
(where the effects of scattering-in were not considered). 
As we describe below, however, complete reproduction of the event distribution was not achieved, and will require further studies.

\subsection{Comparison with the GEANT distribution: absence of the fuel}

The primary objective of the test experiment at the JAPC was to verify whether the fuel could be imaged with our system. 
Throughout the experimental periods the fuel was not loaded, therefore, direct verification was not possible. 
Instead of relying on the GEANT simulations, the images near the fuel loading zone were used to calibrate the 
GEANT distribution and we then extend this understanding to the fuel region.
For the GEANT simulation, three typical cases were considered:
\begin{enumerate} 
\item{The pressure vessel RPV was empty (no water and no fuel). 
Since no material (such as fuel brackets) were considered, this situation is equivalent to completely empty.}
\item{The RPV was filled with water. 
Since material other than the fuel material can be approximated by water, 
we expect to observe distributions obtained in this situation.}
\item{A 4~m cubic fuel source was loaded. 
The fuel material was assumed to be UO$_2$, with an average density of 2.5~g/cc \cite{fuel}.
The rest of the RPV space was filled with water. }
\end{enumerate}

\subsubsection{Vertical distribution}

The vertical event yield distribution around the fuel loading zone was investigated at first.
The GEANT distributions as obtained were limited in describing the detailed structures of the plant, showing a maximum discrepancy of about 10\% in the shape in the vertical range from $-8$~m to $+10$~m about the fuel loading zone center.
 
\begin{figure}[htbp]
\begin{center}
\includegraphics[width=9.0cm]{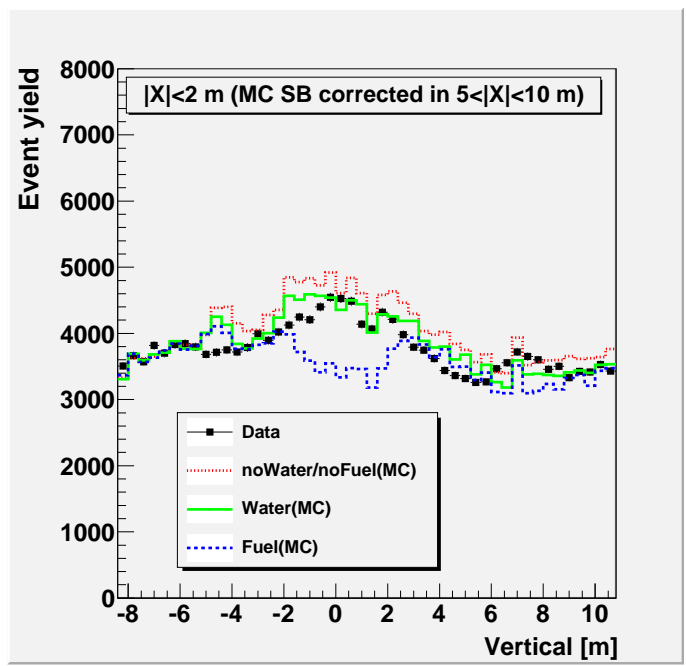}
\end{center}
\caption{Vertical event yield distributions in the range $|X|<2$~m about the fuel loading center,
where the GEANT distributions were corrected for 
sidebands of $X$ ranging from $5<|X|<10$~m. 
The RPV was filled with water (except noWater/noFuel case) in the vertical region from $-5$~m to 11~m. 
The vertical origin is set at the fuel loading zone center. }
\label{fig:vertcorr}
\end{figure}

The incomplete implementation of material conditions and the possibility that an inappropriate muon momentum assumption was used, 
were examined by comparing the experimental data and the GEANT distributions in an area outside of the area under investigation.
Fig.~\ref{fig:vertcorr} shows the vertical event yield distributions in the fuel loading range $|X|<2$~m 
where the GEANT distributions were normalized 
in each vertical slice (0.4~m bin) to the corresponding data using the yields 
in the horizontal sidebands of $X$ over the range $5<|X|<10$~m.
The overall GEANT event yields were normalized to the data in the vertical range $Y<-5$~m 
where the assumption of the water and fuel conditions does not influence on the vertical distribution.
The corrected GEANT distributions described the observed distributions fairly well, 
with the observed data distribution best described by the ``water-only'' assumption (2).

\subsubsection{Horizontal distribution}

\begin{figure}[htbp]
\begin{center}
\includegraphics[width=9.0cm]{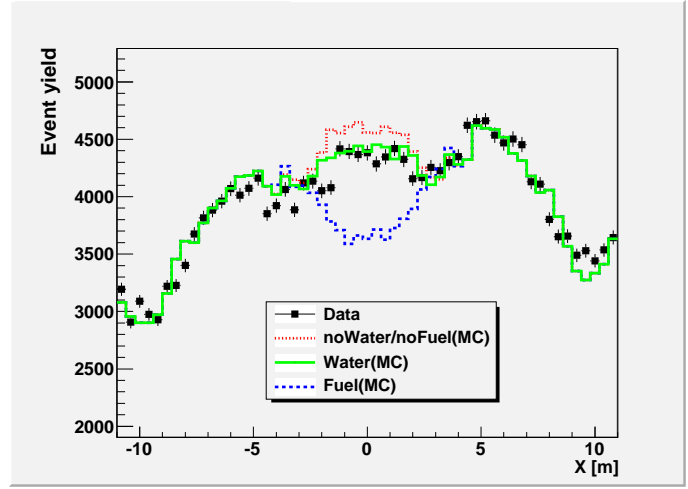}
\end{center}
\caption{Horizontal distributions in the 4~m height range about the fuel loading center. 
The dips at $|X|=10$~m correspond to the containment vessel walls. 
The experimental data (black points) are compared with GEANT simulations under one of three assumptions: 
no water and no fuel in RPV (red dotted); RPV filled entirely with water (green solid); 
and with the fuel loaded with the remaining part filled with water (blue dashed). 
Simulations assumed that the fuel had dimensions of 4~m $\times$ 4~m $\times$ 4~m and an average density of 2.5 g/cc.
}
\label{fig:center}
\end{figure}

Fig.~\ref{fig:center} shows the horizontal event yield distributions for the 4~m height range corresponding to the height of the fuel when loaded. 
The data distributions are compared to the three cases employed in the GEANT simulations. 
A constant value (about 10\% of the event yield) was added to the simulation data to reproduce the structure in the range 
from the containment vessel wall to outside of the RPV. 
As shown, the characteristic distribution around the containment vessel was well reproduced by the simulations, 
which is encouraging with respect to interpolation of the simulation distribution to inside the fuel loading zone.  
The data points inside the RPV were in good agreement with the prediction 
where the RPV was fully with water, and can exclude the other two cases.

\begin{figure}[htbp]
\begin{center}
\includegraphics[width=8.0cm, bb=0 0 523 576]{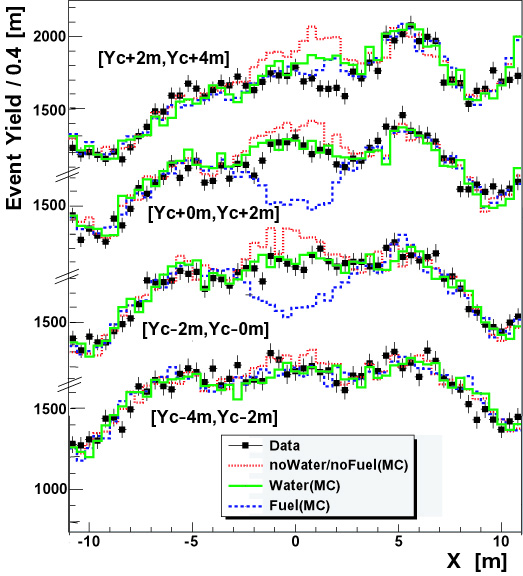}
\end{center}
\caption{Horizontal distributions over a 2~m height range above and below the fuel loading zone. 
The height ranges are shown with respect to the central height of the fuel loading zone, 
{\it e.g.}\,, [Yc--4m,Yc--2m] refers to a range from 4 to 2~m below the fuel loading center.
}
\label{fig:distother}
\end{figure}

Fig.~\ref{fig:distother} shows similar distributions in other Y slices measured across 2~m bands. 
The simulated distributions were generally in agreement with the experimental data 
thereby concluding that this method can reliably confirm the absence of the fuel in the fuel loading zone. 
Note that the observed events in [Yc$+$2m,Yc$+$4m] exhibit a systematical reduction around $X=$2~m. 
This is suspected to be caused by un-accounted reactor equipment located behind on the floor.
The detail of the material is not informed and hence not implemented in the simulation.

\subsection{Sensitivity evaluated using the GEANT simulations}

\begin{figure}[htbp]
\begin{center}
\includegraphics[width=8.0cm]{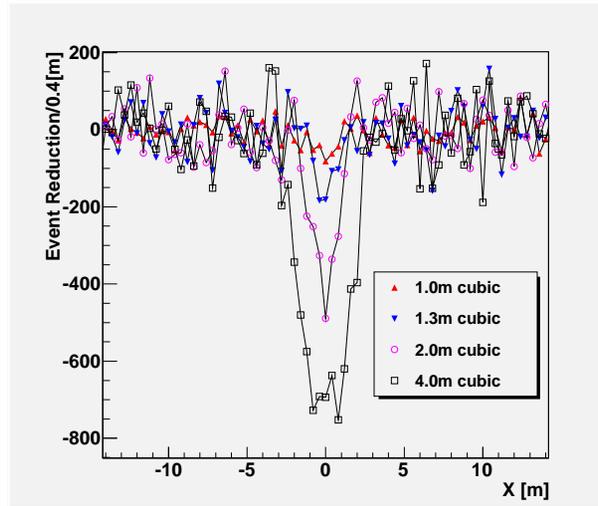}
\end{center}
\caption{Horizontal yield distributions for a 1.6~m Y-slice for different fuel sizes, with 
the distribution obtained for the water only assumption subtracted. 
The fuel was assumed to be cubic with a side length varying from 1~m to 4~m and a density of 2.5 g/cc.} 
\label{fig:days_dist}
\end{figure}

As demonstrated in Fig.~\ref{fig:center}, the GEANT simulations can describe the observed data around the fuel region after addition of a constant. 
The simulation study was extended to investigate the sensitivity of this technique for smaller fuel sizes.
Fig.~\ref{fig:days_dist} shows the horizontal distributions for different fuel sizes: 1~m  to 4~m (cubic), with the distribution for 
the ``water-only'' assumption subtracted. 
In the figure, we assumed the same data taking configuration as PERIOD-1 (114 days, and at a distance of 64~m) 
with the $y$-axis representing the number of events reduced by the existence of fuel material 
in the fuel loading zone center. 
The vertical slice range was fixed at 1.6~m throughout.

The observation sensitivity $S$ is expressed as:

\begin{equation}\notag
\displaystyle S=\frac{[ N_{\rm MC}(0) - N_{\rm MC}(w)]}{\sqrt{N_{\rm obs}}}
\end{equation}

\noindent  
where $N_{\rm obs}$ is the number of observed events taken from the experimental data of PERIOD-1, 
and $N_{\rm MC}(w)$ is the number of simulated events for each fuel size $w$ (0 refers to water-only). 
By fixing the vertical range to 1.6~m for all cases,
we counted the number of events for different horizontal ranges and calculated the sensitivity $S$. 

\begin{figure}[htbp]
\begin{center}
\includegraphics[width=8.0cm]{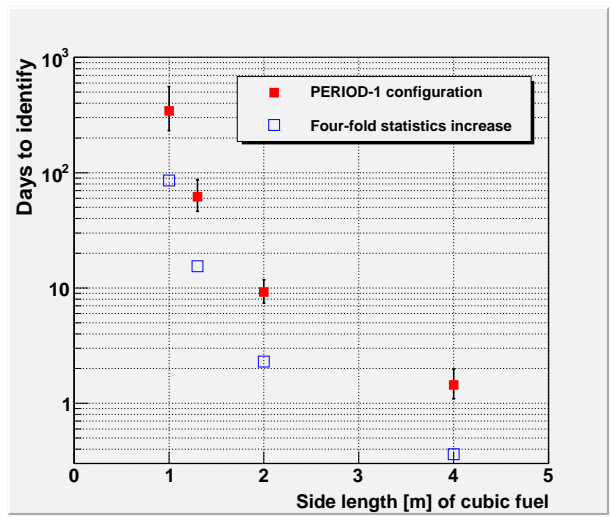}
\end{center}
\caption{Number of days required to confirm the existence of the fuel as a function of the side length of the cubic fuel. 
Data shows the measurement conditions at the JAPC for 64 m from the reactor center with one detector system (filled squares), 
and four-fold increase in the statistics achieved either placing the detector 31 m from the reactor center, or operating 
four systems simultaneously (open squares). }
\label{fig:days}
\end{figure}

We assumed that the existence of the fuel can be confirmed for $S>5$ and  
the number of days to achieve this condition is shown in Fig.~\ref{fig:days} as a function of the fuel size. 
The quoted uncertainties were due to the variation in the side band normalization.
Observations of 2~m (cubic), or larger fuel objects should be achievable in two weeks. 
Also shown in the figure are the results for the case of four-fold increase in event statistics 
that is achievable should the system be placed closer (32~m) or should four detector  
systems be operated simultaneously at a distance of 64~m. 
In this configuration, a 1.3~m (1~m) fuel object should be 
identifiable over a one (three to four) month period.

\subsubsection{Estimation of the fuels in the storage pool}

In PERIOD-2, the detector was directed towards the fuel storage pool. 
The horizontal event yield distribution over a 4~m band across the pool is compared here to simulations. 
In the simulations, the fuel object was assumed to be arranged in one rectangular block 
4~m high, 10~m wide, and 0, 4, or 8 m deep. 
The height was a standard value for the fuel object. 
The width was set wide enough so that the data distribution in 
the corresponding area was not represented by the depth$=$0~m distribution.

\begin{figure}[htbp]
\begin{center}
\includegraphics[width=12.0cm, bb=0 0 710 333]{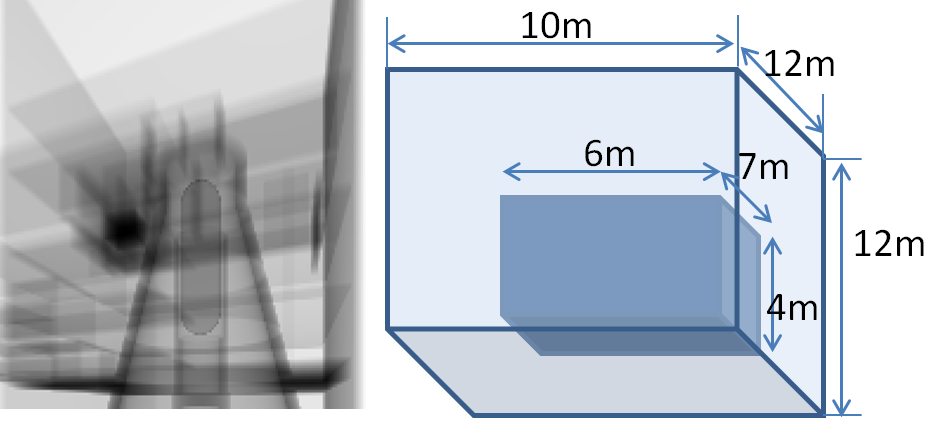}
\includegraphics[width=8.0cm]{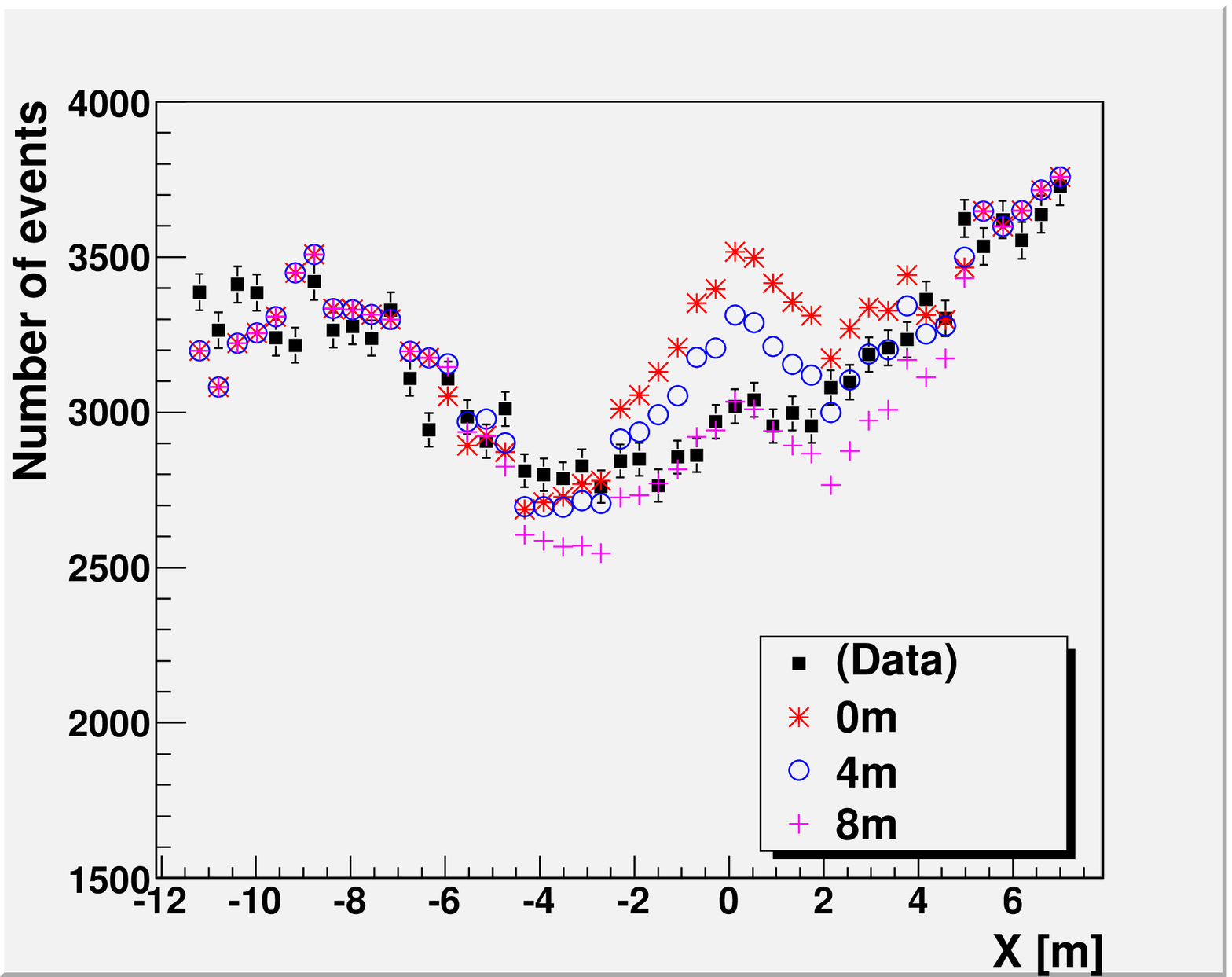}
\includegraphics[width=8.0cm]{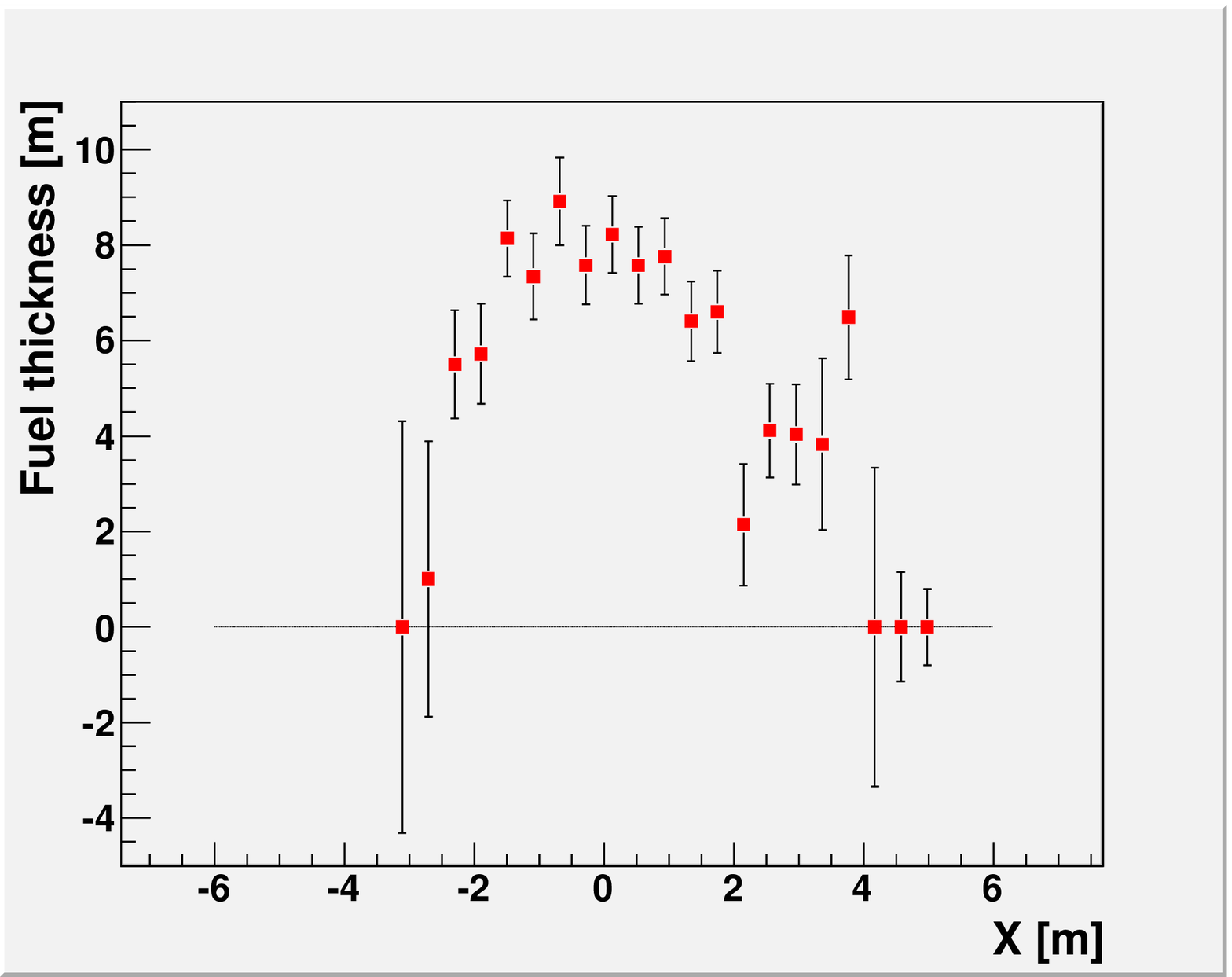}
\end{center}
\caption{(Top) Estimated profile of the fuel object in the storage pool \cite{plant}.
 (Middle) The event yield distributions in a 4~m vertical band across the fuel storage pool (closed square) 
compared to simulations where the fuel object thickness was assumed to be 8 m (plus), 4 m (circle), or 0~m (star). 
(Bottom) The fuel thickness was evaluated as function of the horizontal position (where the density was assumed to be 2.5 g/cc). 
The horizontal origin is set at the center of the pool.
} 
\label{fig:fuelpool}
\end{figure}

The data taken over 99 days, and the simulation distributions are compared in Fig.~\ref{fig:fuelpool}. 
The fuel thickness as a function of the horizontal position was estimated by linear interpolation of 
the simulation distributions for the three thicknesses. 
As shown in the figure, the estimated fuel spans horizontally about 6~m with a typical thickness of 7~m (there may be a thinner part on the right).
The present estimate of the thickness relies on the chosen fuel density of 2.5 g/cc, 
hence the actual thickness will vary inversely with respect to the actual average density.

\section{Conclusion}

We have constructed a muon radiography system consisting of 1 $\times$ 1~m tracking planes utilizing 1~cm wide scintillator bars. 
The system performance was evaluated by identifying the inner structure of the Japan Atomic Power company (JAPC) nuclear plant. 
The detector system was installed 64~m away from the center of the reactor and 
collected 6.3 million muons during PERIOD-1 (114 days), and 7.1 million muons during PERIOD-2 (99 days), whilst targeting the fuel loading zone and the fuel storage pool, respectively. 

The detector can reconstruct an image of the characteristic reactor structures, such as the containment vessel, 
pressure vessel, floors, building walls, and storage pools. 
The image distribution was consistent with the case where the pressure vessel was filled with water or equivalent, 
and no nuclear fuel was identified in the pressure vessel, confirming that the reactor was under maintenance during the data acquisition periods. 
We also measured the profile of the fuel material located in the fuel storage pool.
The data obtained was used to examine the sensitivity of the system. 
The change in water level of the dryer-separator pool and the removal of the well shield plug were clearly identified. 
The spatial resolution was found to be 0.5~m. 
 GEANT simulations tuned to reproduce the observed profiles predicted that a 2~m (cubic) fuel material 
should be able to identifiable within a week. 
By operating multiple systems simultaneously, or moving the system closer to the reactor, 
detection of 1.3~m (cubic) fuel material should also be achievable within a month.

The constructed system operated reliably over a period of seven months,
where an interruption (due to a malfunction of the clock generator) only occurred once.
 The compactness of the detector -- fitting in a shipping container -- is also to be noted. 
The demonstrated ability to investigate the inner structure of the reactor, together with its compactness and reliable operation, 
are very promising features in the adoption of a muon radiography system to image the inner structure
of nuclear plants.

\section*{Acknowledgments}

We would like to thank the Japan Atomic Power Company for permitting us to locate the detector system on their site. 
Without their understanding, we would not have been able to accomplish the present study. 
We wish to particularly acknowledge H. Okuda for informative discussions on conducting the 
measurement at the JAPC whilst adhering to the Nuclear Source Material regulations.
H. Yokomizo of Japan Atomic Energy Agency and H. Sugawara of Okinawa Institute of Science and Technology Graduate University are also to be acknowledged for their support and advice 
for conducting the study at the JAPC. 
The detector system was designed and constructed in a very short period and this 
was realized through the excellent work by the involved companies and 
the help provided by many people of KEK; M. Tanaka for the MPPC DAQ design; 
Y. Yasu for DAQ online programming; and J. Haba for budgetary management.



\end{document}